\documentstyle[prb,aps,multicol]{revtex}
\input epsf

\begin{document}
\draft

\title{   Spin stiffness and quantum fluctuations 
          in $C$-type and $A$-type antiferromagnets     }

\author { Marcin Raczkowski and Andrzej M. Ole\'{s} }
\address{ Marian Smoluchowski Institute of Physics, Jagellonian 
          University, Reymonta 4, PL-30059 Krak\'ow, Poland }

\date{21 March 2002}
\maketitle

\begin{abstract}
We present a systematic study of quantum fluctuations in the $C$-type 
and $A$-type antiferromagnetic (AF) phases in cubic lattices and in 
bilayer systems. Using the linear spin-wave theory, we show that the 
spin stiffness and the quantum corrections to the order parameter and 
energy obtained for $C$-AF and $A$-AF phases decrease with the 
increasing number of ferromagnetic bonds. Therefore, the quantum spin 
effects in LaMnO$_3$ and in LaVO$_3$ are rather small, suggesting the 
magnetic moments of $\sim 3.91$ and $\sim 1.89\mu_B$, respectively. 
They cannot explain the strong reduction of the magnetic order 
parameter observed in cubic vanadates. [Published in Phys. Rev. B {\bf 66}, 
094431 (2002)]
\end{abstract}


\begin{multicols}{2}

The undoped transition metal oxides are characterized by large local
Coulomb interactions $\propto U$, which lead to several fascinating 
phenomena such as high-temperature superconductivity and ``colossal 
magnetoresistance''.\cite{Ima98} When the Coulomb interactions dominate 
over the kinetic energy, the charge fluctuations are quenched and the 
magnetic properties follow from the effective low-energy superexchange 
interactions. In some of these systems the orbital degrees of freedom 
play a role due to the partial filling of (almost) degenerate $d$ 
orbitals, the superexchange interactions are strongly frustrated,
\cite{Tok00,Ole01} and the quantum effects are enhanced.\cite{Fei97} 
These interactions together with the Jahn-Teller effect may induce the 
orbital ordering below a structural transition and break the cubic 
symmetry of the perovskite lattice. In such systems, although the 
crystalographic directions in a three-dimensional (3D) cubic lattice 
are {\it a priori\/} equivalent, one finds magnetic interactions not 
only of different value, but even of {\it different sign\/}, stabilizing 
$C$-type or $A$-type antiferromagnetic (AF) phases.\cite{Ole01} 

One of the best known examples of the non-cubic magnetic interactions in
a perovskite system is the $A$-type AF order observed in LaMnO$_3$,
\cite{Wol55} with ferromagnetic (FM) superexchange within $(a,b)$ planes 
($J_{ab}$) and AF interactions along the $c$ axis ($J_c$), or in 
KCuF$_3$, an almost perfect one-dimensional (1D) Heisenberg 
antiferromagnet.\cite{Ten91} In both above cases the magnetic ordering 
is supported by the orbital ordering which is induced either by 
the Jahn-Teller effect,\cite{Hot99,Su00} or by the superexchange 
interactions.\cite{Fei99,Oka02} Recently it was suggested that the 
latter contribution dominates,\cite{Oka02} but this issue is still 
controversial and has to be clarified by future studies. The spin waves 
in LaMnO$_3$ have been investigated in great detail and it was 
established that the AF interactions $J_c$ are weaker than the FM 
$J_{ab}$ ones,\cite{Fei99} in good agreement with the experimental 
data.\cite{Mou96}

An inverse situation with respect to manganites and cuprates, with AF 
interactions within $(a,b)$ planes coexisting with FM superexchange 
along the $c$-axis, is encountered in the so-called $C$-AF phase, 
observed in cubic vanadates: in LaVO$_3$ below the N\'eel temperature 
$T_N\simeq 143$ K,\cite{Miy00} and in YVO$_3$ at intermediate 
temperatures $77<T<116$ K.\cite{Ren00} Finally, $G$-type AF order, with 
cubic symmetry and the magnetic order parameter staggered in all three 
directions, is found in CaMnO$_3$, LaTiO$_3$, and also in the 
low-temperature phase of YVO$_3$.\cite{Ren00} Particularly this last 
example shows that the type of magnetic order observed in transition 
metal oxides may be triggered by a delicate balance of magnetic 
interactions induced by the orbital ordering.

Other examples of non-cubic antiferromagnets are found in bilayer 
systems, where the effective dimensionality is reduced and the cubic 
symmetry is explicitly broken by geometry, even when the exchange 
interactions are AF and identical on the bonds in different 
crystallographic directions. The $G$-AF structure is realized in 
YBa$_2$Cu$_3$O$_{6+x}$,\cite{Kas98} while the bilayer manganites 
La$_{2-2x}$Sr$_{1+2x}$Mn$_2$O$_7$ show interesting dependence of the 
magnetic order on the doping, with FM structure for $x\simeq 0.4$ (Ref. 
\onlinecite{Per01}) and $A$-AF structure for $x\simeq 0.48$,\cite{Koi01} 
separated by a $C$-AF phase at intermediate doping.\cite{Chu01}

The renewed interest in the magnetic properties of transition metal 
oxides motivates a systematic study of quantum fluctuations in different 
AF structures. The spin-wave theory was introduced long ago,\cite{And52} 
and high accuracy results of the $1/S$ expansion were presented for a 
two-dimensional (2D) Heisenberg antiferromagnet.\cite{Tak89,Cas91} More
recently the properties of 2D dimerized models were studied.\cite{Sir02}
However, we are not aware of any systematic investigation of the spin 
stiffness and the quantum effects in non-cubic systems with different 
strengths and signs of exchange interactions. In this paper we consider 
the quantum fluctuations in the AF phases realized in transition metal 
perovskites, and compare them with those known for the 2D square and 
3D cubic lattice. Further motivation comes from the experimental data 
for numerous systems which could be properly understood only when spin
effects would be extracted from the full quantum problem. Therefore, we 
neglect the orbital fluctuations, and study the spin quantum effects 
alone in the structures stabilized by the superexchange fixed by a 
rigid orbital background. 

We consider the (effective) 3D Heisenberg model with nearest-neighbor 
superexchange interactions, $J_{ab}$ in $(a,b)$ planes and $J_c$ along
$c$ axis, between spins $S$ in $G$-AF, $C$-AF and $A$-AF phases, given 
by
\begin{equation}
{\cal H}_{3D}=
J_{ab}\sum_{\langle ij\rangle\parallel (a,b)}{\bf S}_i\cdot{\bf S}_j
  +J_c\sum_{\langle ij\rangle\parallel     c}{\bf S}_i\cdot{\bf S}_j.
\label{eq:h}
\end{equation}
In cubic crystals the $G$-AF phase is obtained for $J_{ab}>0$ and 
$J_c>0$, while FM interactions stabilize either $C$-AF phase if 
$J_c<0$, or the $A$-AF phase if $J_{ab}<0$. Similar phases are 
possible in the bilayer structures, with the second sum involving only 
the interlayer bonds. We will investigate the range of parameters with 
the interlayer coupling $|J_c|\le 2J_{ab}$, where the ground state is 
ordered. In contrast, for $J_c/J_{ab}>2.55$ the magnetic properties 
are dominated by the singlets which form on interlayer bonds $\langle 
ij\rangle\parallel c$ axis, the long-range order is lost, 
and the spin-wave expansion does not apply.\cite{Mil94} As a reference 
system, we also performed calculations for the 2D square lattice with 
different interactions along the bonds parallel to $a$ and $b$ axis, 
respectively, described by,
\begin{equation}
{\cal H}_{2D}=
 J_a\sum_{\langle ij\rangle\parallel a}{\bf S}_i\cdot{\bf S}_j
+J_b\sum_{\langle ij\rangle\parallel b}{\bf S}_i\cdot{\bf S}_j.
\label{eq:h2d}
\end{equation}
We assume $J_a>0$ and study both the $G$-AF ($J_b>0$) and $C$-AF 
($J_b<0$) phase.

We determined the dispersion of spin waves in the spin models given by
Eqs. (\ref{eq:h}) and (\ref{eq:h2d}), the quantum corrections to the 
magnetic order parameter $\langle S^z\rangle$, and to the ground state 
energy $E=\langle{\cal H}_{3D}\rangle$ ($E=\langle{\cal H}_{2D}\rangle$) 
using the linear spin-wave theory (LSWT) in the leading $1/S$ order.
First we transform the spin operators at the sites $j\in B$ sublattice:
$S_j^{\pm}\to S_j^{\mp}$, $S_j^z\to -S_j^z$, which removes spins down 
$\langle S_j^z\rangle_0=-S$ of the N\'eel state. Next we introduce 
bosonic operators $\{a_i^{\dagger},a_i^{}\}$, and use the lowest order 
of the Holstein-Primakoff transformation:\cite{Mat81} 
\begin{equation}
S_i^z=S-a_i^{\dagger}a_i^{}, \hskip .5cm
S_i^+\simeq \sqrt{2S}a_i^{}, \hskip .5cm
S_i^-\simeq \sqrt{2S}a_i^{\dagger}. 
\label{eq:hp}
\end{equation}
The excitations are derived from the equations of motion for the 
energy-dependent Green's functions,\cite{Zub60,Hal72} 
\begin{equation}
E\langle\langle a_i^{}|a_j^{\dagger}\rangle\rangle=
      {1\over 2\pi}\delta_{ij}
+ \langle\langle [a_i^{},{\cal H}_{3D}]|a_j^{\dagger}\rangle\rangle. 
\label{eq:motion}
\end{equation}
After the Fourier transformation the spin waves are found by a 
Bogoliubov transformation which diagonalizes a $2\times 2$ dynamical 
matrix at each {\bf k} point. The positive magnon energies for the 3D 
$G$-AF, $C$-AF and $A$-AF phase are:
\begin{eqnarray}
\label{eq:omega_g}
\omega_G({\bf k})&=& 2S\{(2J_{ab} + J_c)^2                 \nonumber \\
    & &\hskip .5cm  -[2J_{ab}\gamma_+({\bf k}) + 
    J_c\gamma_z({\bf k})]^2\}^{1/2},                                 \\
\label{eq:omega_c}
\omega_C({\bf k})&=& 2S\{[2J_{ab}+|J_c|(1-\gamma_z({\bf k}))]^2 
                                                           \nonumber \\
    & &\hskip .5cm  -4J_{ab}^2\gamma_+^2({\bf k})\}^{1/2},           \\
\label{eq:omega_a}
\omega_A({\bf k})&=& 2S\{[2|J_{ab}|(1-\gamma_+({\bf k}))+J_c]^2
                                                           \nonumber \\
    & &\hskip .5cm  -   J_c^2 \gamma_z^2({\bf k})\}^{1/2}, 
\end{eqnarray}
where $\gamma_+({\bf k})=\case{1}{2}\left(\cos k_x+\cos k_y\right)$ and 
$\gamma_z({\bf k})=\cos k_z$ are {\bf k}-dependent structure factors.
The corresponding magnon energies for the bilayer systems are:
\begin{eqnarray}
\label{eq:omega_gbi}
\omega_G^{\parallel}({\bf k}) &=& S\{(4J_{ab} + J_c)^2 
      -[4J_{ab}\gamma_+({\bf k}) + \lambda J_c]^2\}^{1/2},           \\
\label{eq:omega_cbi}
\omega_C^{\parallel}({\bf k}) &=& S\{[4J_{ab} + |J_c|
    (1-\lambda)]^2-16J_{ab}^2\gamma_+^2({\bf k})\}^{1/2},            \\
\label{eq:omega_abi}
\omega_A^{\parallel}({\bf k}) &=& S\{[4|J_{ab}|
    (1-\gamma_+({\bf k}))+J_c]^2-\lambda^2J_c^2\}^{1/2}, 
\end{eqnarray}
where $\lambda=\pm 1$ corresponds to $k_z=0,\pi$, i.e., to the symmetric
and antisymmetric interlayer modes, respectively. The magnon energies 
for the $G$-AF and $C$-AF phase in the 2D case are:
\begin{eqnarray}
\label{eq:omega_g2D}
\omega_G^{2D}({\bf k}) &=& 2S\{(J_a \!+\! J_b)^2             
  \!+\! [J_a\gamma_x({\bf k}) 
                \!+\! J_b\gamma_y({\bf k})]^2\}^{1/2},               \\
\label{eq:omega_c2D}
\omega_C^{2D}({\bf k}) &=& 2S\{[J_a\!+\!|J_b|(1-\gamma_y({\bf k}))]^2 
  \!-\!J_a^2\gamma_x^2({\bf k})\}^{1/2},
\end{eqnarray}
with $\gamma_x({\bf k})=\cos k_x$, and $\gamma_y({\bf k})=\cos k_y$.

The size of quantum fluctuation corrections to the classical order 
parameter $\langle S_i^z\rangle_0=S$, and the intersite spin fluctuations
$\propto \langle S_i^-S_j^+\rangle$ which modify the energy, determine 
the stability of the classical phases. One finds that the order parameter 
for $i\in A$ sublattice is reduced by local quantum fluctuations,
\begin{equation}
\langle S_i^z\rangle
= S -       \langle S_i^-S_i^+\rangle
= S - \delta\langle        S^z\rangle,
\label{eq:order}
\end{equation}
and the local correlation function $\langle S_i^-S_i^+\rangle$ is 
determined by the Green's function 
$\langle\langle a_{\bf k}^{}|a_{\bf k}^{\dagger}\rangle\rangle$ by,
\cite{Hal72}
\begin{equation}
\langle S_i^-S_i^+\rangle={1\over N}\sum_{\bf k}
\int_{-\infty}^{+\infty}d\omega\;
{\cal I}({\bf k},\omega)\;{1\over \exp(\beta\omega)-1},
\label{eq:BA1}
\end{equation}
where $\beta=1/k_BT$, $N$ is the number of sites, and
\begin{eqnarray}
{\cal I}({\bf k},\omega) &=& 2\;{\rm Im}\langle\langle 
a_{\bf k}^{}|a_{\bf k}^{\dagger}\rangle\rangle_{\omega-i\epsilon}
                                             \nonumber\\
&=& \sum_{\nu}{\cal A}^{(\nu)}({\bf k})\;\delta
(\omega -  \omega _{\bf k}^{(\nu)})
\label{eq:density}
\end{eqnarray}
is the spectral density of spin excitations, with the operator
$a_{\bf k}^{}$ being a Fourier tranform of $a_i^{}$. The quantum 
corrections are found by taking $T\to 0$ limit of Eq. (\ref{eq:BA1}), 
where the averages $\langle S^-_iS^+_i\rangle$ are determined from the 
spectral weights ${\cal A}^{(\nu)}({\bf k})$ in Eq. (\ref{eq:density}) 
of the excitations at negative frequencies, and therefore,
\begin{equation}
\delta \langle S^z\rangle=
{1\over N}\sum_{\bf k}\sum_{\nu <0}{\cal A}^{(\nu)}({\bf k}).
\label{eq:BA2}
\end{equation}
Thus, using Eqs. (\ref{eq:density}) and (\ref{eq:BA2}) one obtains the 
quantum corrections $\delta\langle S^z\rangle$ for the 3D phases:
\begin{eqnarray}
\label{eq:order_g}
\delta\langle S^z\rangle_G &=& S\int{d^3{\bf k}\over (2\pi)^3}
{2J_{ab}+J_c\over \omega_G({\bf k})}-{1\over 2},                    \\ 
\label{eq:order_c}
\delta\langle S^z\rangle_C &=& S\int{d^3{\bf k}\over (2\pi)^3}
{2J_{ab}+|J_c|[1-\gamma_z({\bf k})]\over \omega_C({\bf k})}
-{1\over 2},                                                        \\
\label{eq:order_a}
\delta\langle S^z\rangle_A &=& S\int{d^3{\bf k}\over (2\pi)^3}
{2|J_{ab}|[1-\gamma_+({\bf k})]+J_c\over \omega_{A}({\bf k})}
-{1\over 2},
\end{eqnarray}
with magnon dispersions given by Eqs. 
(\ref{eq:omega_g})--(\ref{eq:omega_a}). The quantum corrections for the 
bilayer structures and for the square lattice can be obtained in a 
similar way -- they involve 2D integrations over the respective
excitation spectra, and, in addition, the summation over the symmetric
and antisymmetric modes for the bilayer system. 

The quantum corrections $\delta E$ to the ground state energy per site, 
$E=E_0-\delta E$, where $E_0$ is the energy found using the classical 
state, were obtained by integrating the excitation spectra 
$\omega_M({\bf k})$, with $M=A,C,G$. For instance, for 3D systems one 
finds $E_0=-(2|J_{ab}|+|J_c|)S^2$, and 
\begin{equation}
\delta E=(2|J_{ab}|+|J_c|)S
        -{1\over 2}\int{d^3{\bf k}\over (2\pi)^3}\;\omega_M({\bf k}).
\label{eq:energy}
\end{equation}
In order to compare the systems of different dimensionality we use 
below the relative energy correction $S(\delta E/E_0)$ as a measure of 
the quantum correction to the ground state energy. For the 3D AF systems 
with $|J_{ab}|=|J_c|=J$ (or for the 2D systems with $|J_a|=|J_b|=J$) the 
relative energy correction $\delta E/E_0$ can be expressed by,
\begin{equation}
\frac{\delta E}{E_0} = \frac{\gamma}{zS}, 
\label{eq:gamma}
\end{equation}
where $z$ is the number of nearest neighbors, and the coefficient 
$\gamma>0$ provides a measure of quantum effects. It depends on the 
system dimensionality, on lattice type, and on the type of AF order. 
Therefore, we write the ground state energy as follows:\cite{Mat81}
\begin{equation}
E = -{1\over 2}zJS^2\left(1 + {\gamma\over zS}\right). 
\label{eq:Mattis}
\end{equation}

Consider first a cubic crystal with equal strength of AF and FM 
superexchange interactions $(|J_{ab}|=|J_c|=J)$. \\
\begin{figure}
\centerline{
\epsfxsize=0.4\textwidth
\epsfbox{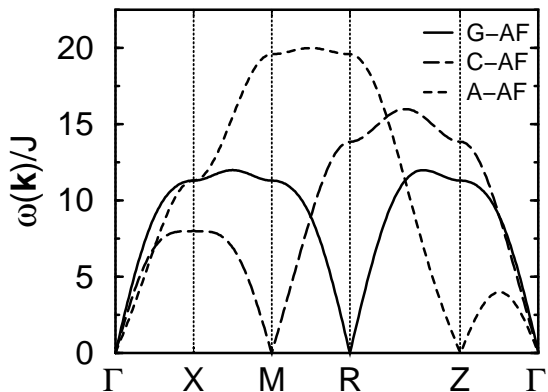}}
\vskip 0.3cm
\caption{
The spin-wave dispersions as obtained within the LSWT in the $G$-AF, 
$C$-AF and $A$-AF phases along the main directions of the 3D BZ for 
a simple cubic lattice, with equal AF and FM exchange interactions 
$(|J_{ab}|=|J_c|=J)$. We have used the standard labels for high-symmetry 
points: $\Gamma=(0,0,0)$, $X=(\pi,0,0)$, $M=(\pi,\pi,0)$, 
$R=(\pi,\pi,\pi)$, and $Z=(0,0,\pi)$.
}
\label{fig:omega}
\end{figure}
\noindent
The spin-wave 
excitations with positive energies are presented in Fig. \ref{fig:omega} 
along the high-symmetry directions of the 3D Brillouin zone (BZ). One
finds an expected Goldstone mode with $\omega_M({\bf k})=0$ at 
$R=(\pi,\pi,\pi)$, $M=(\pi,\pi,0)$, and $Z=(0,0,\pi)$ point for $G$-AF, 
$C$-AF and $A$-AF phase, respectively. It corresponds to the vector in 
the reciprocal space which couples two spin excitations in a given AF 
structure. The spin excitations are linear near the Goldstone mode, 
$\omega_M({\bf k})\simeq Dk$, and close to the $\Gamma$ point in each AF 
structure, and the spin-wave stiffness constant $D$ is {\it isotropic\/}, 
being the same for any cubic direction, in spite of the different signs 
of superexchange constants. Note also that the width of the magnon 
dispersion increases with the increasing number of FM bonds, and would 
be the largest and equal to $24J$ for the cubic FM phase, with equal FM  
exchange interactions in all three directions (not shown). Increasing 
overall dispersion $\omega_M({\bf k})$ in Eq. (\ref{eq:energy}) reduces 
the quantum correction $\delta E$ when the number of FM bonds increases. 

We compare the quantum effects given by Eqs. (\ref{eq:BA2}) and 
(\ref{eq:gamma}) for different systems listed in Table I. The values
obtained for 2D and 3D $G$-AF phases reproduce the known results given 
by Mattis.\cite{Mat81} Both $\delta\langle S^z\rangle$ and $\delta E$ 
are reduced when some of the bonds are FM. When the dimensionality is 
fixed, the quantum corrections decrease in a systematic way with the 
increasing number of FM interactions, from the $G$-AF to $A$-AF order.
\cite{notebi} A somewhat counterintuitive result is that at the same 
time the stiffness constant $D$ is enhanced only by the AF interactions, 
while it {\it decreases\/} with the increasing number of FM bonds for 
all systems: the 3D cubic (see Fig. \ref{fig:omega}), 2D square, as well 
as for the bilayer structure (Table \ref{table1}). Thus, the magnon 
energy increases slower with $k$ near the $\Gamma$ point by a factor 
$1/\sqrt{3}$ in $A$-AF than in $G$-AF phases. Further, for a given 
magnetic structure the quantum corrections $\delta\langle S^z\rangle$ 
and $S(\delta E/E_0)$ gradually {\it decrease\/} when the dimensionality 
of the system {\it increases\/} from a 2D square lattice thoughout a 
bilayer system to a 3D cubic lattice (Table I). In the bilayer case an isotropic
\begin{table}[b]
\caption{
The quantum corrections to the magnetic order parameter 
$\delta\langle S^z\rangle$ and to the ground state energy $\delta E$, 
and the stiffness constant $D$, as obtained in the LSWT for the phases 
with isotropic low-energy spin excitations $\omega_M({\bf k})\simeq Dk$,
with $M=G,C,A$. Parameters: $|J_{ab}|=|J_c|=J$ for the 2D and 3D AF 
phases, and $|J_{ab}|=|J_c|/2=J$ in the bilayer systems. 
}
\label{table1}
\vskip .3cm
\begin{tabular}{ccccccc}
 & lattice  & AF phase & $\delta\langle S^z\rangle$ &
                                  $S(\delta E/E_0)$ & $D/2JS$ & \\
\tableline
 &2D square &  $G$-AF  &  0.1966  &  0.1580  &  $\sqrt{2}$  & \\
 &          &  $C$-AF  &  0.1214  &  0.0835  &  $1$         & \\
\tableline
 & bilayer  &  $G$-AF  &  0.1589  &  0.1373  &  $\sqrt{3}$  & \\
 &          &  $C$-AF  &  0.1069  &  0.0743  &  $\sqrt{2}$  & \\
 &          &  $A$-AF  &  0.0859  &  0.0825  &  $1$         & \\
\tableline
 & 3D cubic &  $G$-AF  &  0.0783  &  0.0971  &  $\sqrt{3}$  & \\
 &          &  $C$-AF  &  0.0565  &  0.0662  &  $\sqrt{2}$  & \\
 &          &  $A$-AF  &  0.0318  &  0.0340  &  $1$         & \\
\end{tabular}
\end{table}
\begin{figure}
\centerline{
\epsfxsize=0.4\textwidth
\epsfbox{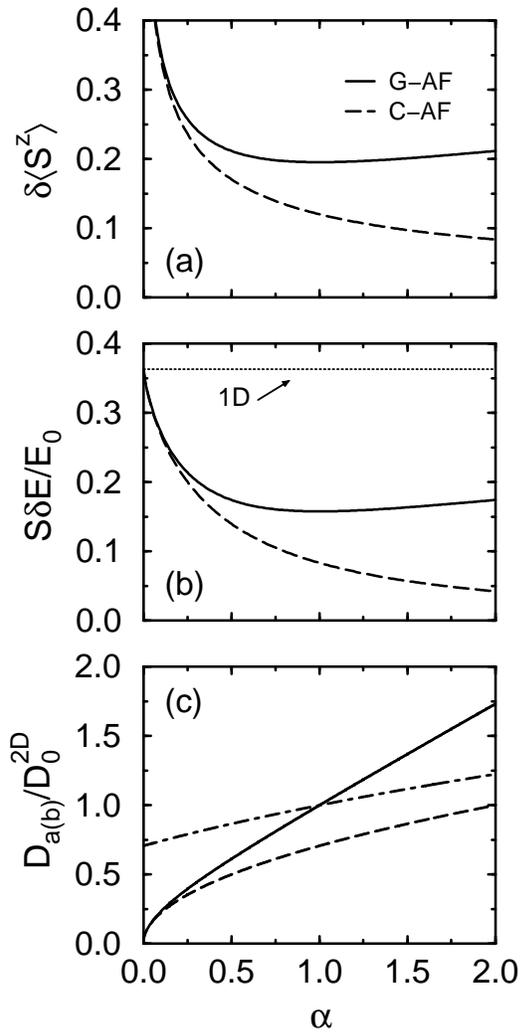}}
\vskip 0.3cm
\caption{
Quantum corrections to: 
(a) the magnetic order parameter $\delta\langle S^z\rangle$, and 
(b) energy $S(\delta E/E_0)$, obtained in the LSWT as functions of 
$\alpha=|J_b|/J_a$ for the $G$-AF and $C$-AF 2D phase. Part (c) shows
the stiffness constants: $D_b$ and $D_a$ for $G$-AF phase (solid and 
dot-dashed line), and $D_b$ for $C$-AF (long-dashed line), normalized 
to the value $D_0^{2D}=2\sqrt{2}JS$ obtained for the $G$-AF phase with
$J_a=J_b$. The LSWT result for $S(\delta E/E_0)$ in the 1D model is 
shown in (b) by dotted line.
}
\label{fig:2D}
\end{figure}
 
\noindent
stiffness constant is obtained only when the
interlayer superexchange interaction is larger by a factor of two than 
the respective intralayer interactions, i.e., $|J_{ab}|=|J_c|/2=J$, 
which simulates the missing second neighbor of each atom along the $c$ 
axis.

The long-range order, with $\langle S^z\rangle>0$ ($\delta\langle 
S^z\rangle<0.5$ for $S=1/2$) [see Fig. \ref{fig:2D}(a)], is stabilized 
within the LSWT in quasi-1D systems realized on a 2D square lattice 
already by relatively small interchain couplings, $|J_b|/J_a\simeq 
0.15$. Qualitatively the values of $\delta\langle S^z\rangle$ and 
$\delta E/E_0$ [Fig. \ref{fig:2D}(b)] behave in a similar way; they 
both decrease for the $C$-AF phase, while the quantum corrections reach 
a minimum for an isotropic 2D antiferromagnet at $J_b=J_a$ in the 
\begin{figure}
\centerline{
\epsfxsize=0.4\textwidth
\epsfbox{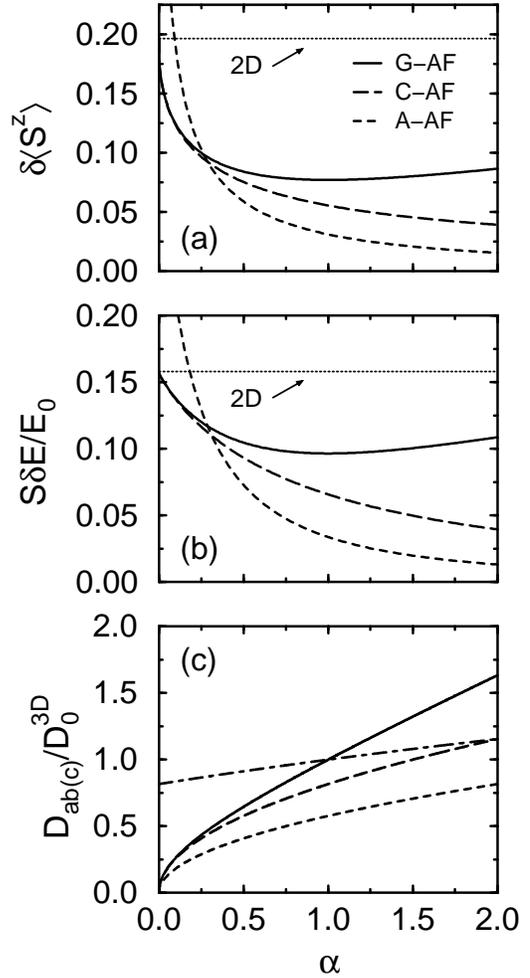}}
\vskip 0.3cm
\caption{
Quantum corrections as in Fig. \ref{fig:2D}, but for the 3D AF phases,
and with $\alpha=|J_c|/J_{ab}$ ($\alpha=|J_{ab}|/J_c$) for the $G$- and
$C$-AF phase ($A$-AF phase). Dotted lines show the values of 
$\delta\langle S^z\rangle$ (a) and $S\delta E/E_0$ (b), obtained within 
the LSWT for the isotropic 2D square lattice. Part (c) shows the 
stiffness constants $D_{\gamma}$ normalized to $D_0^{3D}=2\sqrt{3}JS$ 
obtained for the $G$-AF phase with cubic symmetry of exchange 
interactions: $D_c$ and $D_{ab}$ (solid and dot-dashed line) for 
$G$-AF phase; $D_c$ for $C$-AF phase (long-dashed line), and $D_{ab}$ 
for $A$-AF phase (dashed line). 
}
\label{fig:3D}
\end{figure}
\noindent
$G$-AF phase, and next increase again for $\alpha>1$ when a stronger AF 
interaction occurs along the $b$ axis. 

The magnon stiffness increases when the (either AF or FM) interactions 
along the $b$ axis $|J_b|$ increase, and one finds 
$D_b/D_0^{2D}\simeq\sqrt{\alpha/2}$ for small $\alpha=|J_b|/J_a$ [Fig. 
\ref{fig:2D}(c)], with $D_0^{2D}=2\sqrt{2}JS$ obtained in a $G$-AF phase
with cubic symmetry. However, the behavior of the $G$-AF and $C$-AF 
phase is qualitatively different. First of all, the value of $D_a$ 
increases with $J_b$ in the $G$-AF phase, while it remains constant 
($D_a/D_0^{2D}=1/\sqrt{2}$) in the $C$-AF phase. Second, except for the 
asymptotic regime of $\alpha<0.3$, the values of $D_b$ are considerably 
lower in the $C$-AF phase 
\begin{figure}
\centerline{
\epsfxsize=0.4\textwidth
\epsfbox{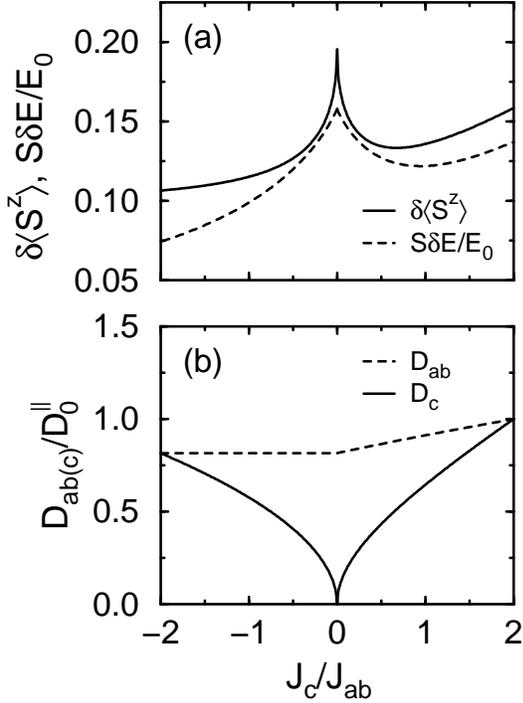}}
\vskip 0.3cm
\caption{
Quantum corrections (a) to the magnetic order parameter
$\delta\langle S^z\rangle$ (solid line), and to the ground state energy 
$S\delta E/E_0$ (dashed line), and (b) the stiffness constants $D_c$ 
and $D_{ab}$ (solid and dashed line) for an AF bilayer system with 
$J_{ab}>0$, as functions of the interlayer coupling $J_c/J_{ab}$, in 
units of $D_0^{\parallel}=2\sqrt{3}JS$ (see Table \protect\ref{table1}).
}
\label{fig:bila}
\end{figure}
\noindent
than those in the $G$-AF phase.

Next we consider $C$-AF and $A$-AF order in 3D systems with increasing 
FM interactions and compare them with the $G$-AF phase (Fig. 
\ref{fig:3D}). Both $\delta\langle S^z\rangle$ and $\delta E/E_0$ 
decrease rapidly in $G$-AF phase with increasing $\alpha=J_c/J_{ab}$ in 
the regime of $0<\alpha<0.5$, while for $\alpha>0.5$ the quantum effects 
are reduced stronger by the FM bonds which make the magnetic order in 
the $C$-AF phase more classical than in the $G$-AF phase. The values of 
$\delta\langle S^z\rangle$ and $\delta E/E_0$ increase again for 
$\alpha>1$ in the $G$-AF phase as the system starts to approach a 
quasi-1D regime with enhanced quantum fluctuations. In the $A$-AF phase 
the strongest quantum fluctuations are obtained for small 
$\alpha=|J_{ab}|/J_c$, where the AF chains along $c$ axis are weakly 
coupled. However, for $\alpha>0.3$ the quantum corrections are already 
smaller than those in the $C$-AF phase, and further decrease with 
increasing $\alpha=J_c/J_{ab}$, i.e., when the FM interactions along the 
$c$ axis get stronger.  

The stiffness constants $D_{ab}$ and $D_c$ shown in Fig. \ref{fig:3D}(c) 
increase with increasing $\alpha$ in all cases, proving that the 
increasing exchange interactions cause magnons to harden. For weak 
interactions along the $c$ axis one finds $D_c/D_0^{3D}\simeq
\sqrt{2\alpha/3}$ for $G$- and $C$-AF phase, and $D_{ab}/D_0^{3D}\simeq
\sqrt{\alpha/3}$ for $A$-AF phase, where $D_0^{3D}=2\sqrt{3}JS$ is 
obtained in a $G$-AF phase with cubic symmetry. We note that both $D_c$ 
and $D_{ab}$ increase when $J_c$ is increased in the $G$-AF phase. In
contrast, the stiffness constants: $D_{ab}$ in the $C$-AF phase, and 
$D_c$ in the $A$-AF phase, are not influenced by the increasing FM 
interactions ($J_c$ and $J_{ab}$, respectively), and one finds:
$D_{ab}/D_0^{3D}=\sqrt{2/3}$ and $D_c/D_0^{3D}=1/\sqrt{3}$ in the $C$-AF 
and $A$-AF phase. This illustrates a general rule -- when the AF 
interactions increase, the spin stiffness increases both along the AF 
and FM bonds. In contrast, the spins on the FM bonds cannot fluctuate 
and thus the changes of FM interactions do not modify the spin stiffness 
along the AF bonds.

The dependence of the quantum corrections and the stiffness constants on 
the interlayer coupling $\propto J_c$ in the bilayer system is 
summarized in Fig. \ref{fig:bila}. As in the 3D case (Fig. 
\ref{fig:3D}), the increasing interlayer coupling gives a fast crossover 
from a system of two independent planes to a bilayer system with reduced 
values of $\delta\langle S^z\rangle$ and $\delta E/E_0$ [Fig. 
\ref{fig:bila}(a)]. It is quite remarkable that the effect of finite and 
small $J_c$ is almost identical for the AF and FM interaction in the 
regime of $|J_c/J_{ab}|<0.3$, and the quantum corrections decrease 
simply due to the dimensional crossover. Only at $|J_c/J_{ab}|>0.3$ the 
$G$-AF and $C$-AF phase start to differ: the quantum corrections pass 
through a minimum and increase again due to the AF interlayer coupling
($J_c>0$) which favors singlet states on these bonds in the $G$-AF 
phase, while they steadily decrease when the FM interlayer coupling 
($J_c<0$) gets stronger in the $C$-AF phase. Similar as in the 3D case, 
$D_c/D_0^{\parallel}\simeq\sqrt{\alpha/3}$ in the regime of small 
$\alpha=|J_c|/J_{ab}<0.3$, where $D_0^{\parallel}=2\sqrt{3}JS$, found
for the $G$-AF phase at $J_c=2J_{ab}$, is used as a unit. In the $C$-AF 
phase the value of $D_{ab}=\sqrt{2/3}D_0^{\parallel}$ is independent of 
$J_c$, while it approaches $D_{ab}=D_0^{\parallel}$ in the  
$G$-AF phase when $|J_c|\to 2J_{ab}$. Finally, as expected, the quantum 
corrections in the $A$-AF bilayer phase increase with increasing AF 
interlayer coupling $J_c>0$, and become even larger than those in the 
$C$-AF phase for sufficiently large $J_c$. One finds equal reduction of 
the order parameter $\delta\langle S^z\rangle\simeq 0.095$ in both 
phases at $|J_c/J_{ab}|\simeq 2.46$.

The present study shows that the spin quantum corrections due to spin
fluctuations are small both in the $A$-AF phase of LaMnO$_3$, and in the
$C$-AF phases of LaVO$_3$ and YVO$_3$. Indeed, large magnetic moments
$\sim 3.87\mu_B$ measured in LaMnO$_3$ (see Ref. \onlinecite{Mou96}) are 
almost perfectly reproduced by the present calculation which gives the 
moment $\langle M\rangle\simeq 2\mu_B\langle S^z\rangle\simeq 3.91\mu_B$ 
for spin $S=2$ and for the experimental ratio of $|J_{ab}|/J_c=1.43$. 
In contrast, the strong reduction of the order parameter in LaVO$_3$ (to 
$\sim 1.3\mu_B$\cite{Miy00}) and YVO$_3$ (to $\sim 1.0\mu_B$ at finite 
temperature $T\simeq 85$ K\cite{Ren00}) cannot be explained by rather 
weak quantum spin effects, reducing the order parameter only down to 
$\sim 1.89\mu_B$ for spin $S=1$ and $|J_c|/J_{ab}\simeq 1$ at $T=0$.
This result indicates that the quantum effects due to {\it orbital 
fluctuations\/} are stronger and dominate the behavior of the $t_{2g}$ 
systems with degenerate orbitals.\cite{Kha01}

Summarizing, we have shown that the quantum corrections to the order
parameter and to the ground state energy {\it decrease\/} systematically 
with the {\it increasing\/} number and strength of FM bonds, from $G$-AF 
through $C$-AF to $A$-AF phase, both in the 3D cubic lattice and in the 
bilayer system. This shows that the AF interactions {\it dominate\/} in 
the non-cubic antiferromagnets. Indeed, when the interactions are AF 
along only {\it one or two\/} cubic directions, their increase causes
the increase of the spin stiffness and of the quantum fluctuations along 
{\it all three\/} cubic directions. 

%%%%%%%%%%%%%%%%%%%%%%%%%%%%%%%%%%%%%%%%%%%%%%%%%%%%%%%%%%%%%%%%%%%%%%%%
%%
%%                           ACKNOWLEDGMENTS
%%
%%%%%%%%%%%%%%%%%%%%%%%%%%%%%%%%%%%%%%%%%%%%%%%%%%%%%%%%%%%%%%%%%%%%%%%%
\acknowledgments

This work was financially supported by the Polish State Committee of 
Scientific Research (KBN), Project No. 5~P03B~055~20.

%%%%%%%%%%%%%%%%%%%%%%%%%%%%%%%%%%%%%%%%%%%%%%%%%%%%%%%%%%%%%%%%%%%%%%%%
%
%                           REFERENCES
%
%%%%%%%%%%%%%%%%%%%%%%%%%%%%%%%%%%%%%%%%%%%%%%%%%%%%%%%%%%%%%%%%%%%%%%%%

\end{multicols}


\begin{references}
\bibitem{notebi} An exception here is the bilayer system, where 
                 isotropic magnons are obtained at $|J_c|=2|J_{ab}|$, 
		 and then the value of $\delta E$ is smaller in the 
		 $C$-AF phase than in the $A$-AF phase.
\end{references}
\end{document}